\documentclass[12pt]{article}
\usepackage{mathrsfs}
\makeatletter \@addtoreset{equation}{section} \makeatother
\renewcommand{\theequation}{\thesection.\arabic{equation}}
\newcommand{\ba}{\begin{array}}
\newcommand{\ea}{\end{array}}
\newcommand{\beq}{\begin{equation}}
\newcommand{\eeq}{\end{equation}}
\newcommand{\bea}{\begin{eqnarray}}
\newcommand{\eea}{\end{eqnarray}}

\def\bce{\begin{center}}
\def\ece{\end{center}}

\def\nonu{\nonumber}

\def\pa{\partial}

\def\be{\beta}

\def\de{\delta}

\def\la{\lambda}

\def\eps6{{\displaystyle \mathop{\epsilon}^{6}}{}}
\def\g6{{\displaystyle \mathop{g}^{6}}{}}

\def\nab6{{\displaystyle \mathop{\nabla}^{6}}{}}

\def\0{{\sst{(0)}}}
\def\1{{\sst{(1)}}}
\def\2{{\sst{(2)}}}
\def\3{{\sst{(3)}}}
\def\4{{\sst{(4)}}}
\def\5{{\sst{(5)}}}
\def\6{{\sst{(6)}}}
\def\7{{\sst{(7)}}}
\def\8{{\sst{(8)}}}

\def\ba{\begin{array}}
\def\ea{\end{array}}
\def\beq{\begin{equation}}
\def\eeq{\end{equation}}
\def\be{\begin{equation}}
\def\ee{\end{equation}}

\def\la{\lambda}
\def\eps{\epsilon}

\def\ba{\begin{array}}
\def\ea{\end{array}}
\def\beq{\begin{equation}}
\def\eeq{\end{equation}}
\def\be{\begin{equation}}
\def\ee{\end{equation}}

\def\la{\lambda}
\def\eps{\epsilon}

\def\eps6{{\displaystyle \mathop{\epsilon}^{6}}{}}

\def\nab6{{\displaystyle \mathop{\nabla}^{6}}{}}

\newcommand{\bean}{\begin{eqnarray*}}
\newcommand{\eean}{\end{eqnarray*}}

\begin{document}
\thispagestyle{empty} \addtocounter{page}{-1}
   \begin{flushright}
\end{flushright}

\vspace*{1.3cm}
  
\centerline{ \Large \bf
Toward A Supersymmetric $w_{1+\infty}$ Symmetry
}
\vspace*{0.3cm}
\centerline{ \Large \bf
in the Celestial Conformal Field Theory } 
\vspace*{1.5cm}
\centerline{ {\bf  Changhyun Ahn}
} 
\vspace*{1.0cm} 
\centerline{\it 
 Department of Physics, Kyungpook National University, Taegu
41566, Korea} 
\vspace*{0.5cm}
\centerline{\tt ahn@knu.ac.kr
} 
\vskip2cm

\centerline{\bf Abstract}
\vspace*{0.5cm}

The $w_{1+\infty}$ symmetry algebra appears in the Einstein--Yang--Mills
theory, proposed recently by Strominger.
In this paper, 
we derive
the supersymmetric $w_{1+\infty}$ symmetry by using the known results on the
operator product expansions (OPEs) between the graviton, gravitino,
gluon, and gluino
in the supersymmetric version of the above theory.
We calculate the four additional commutator relations
between the soft currents explicitly.
In addition, we analyze the works of Odake et al. and Pope et al. and introduce the additional symmetry current
that corresponds to the celestial gluino operator. Through this procedure,
all seven commutator relations
can be connected to the ones associated with
the supersymmetric $w_{1+\infty}$ algebra with $SU(N)$ symmetry
under the
restrictions of wedge modes.


\baselineskip=18pt
\newpage
\renewcommand{\theequation}
{\arabic{section}\mbox{.}\arabic{equation}}

\tableofcontents

\section{ Introduction}

In the tree level Einstein-Yang-Mills theory, the leading operator
product expansions (OPEs) on the celestial sphere of conformal primary
gravitons and gluons are determined \cite{PRSY}.
The structure constants in the right-hand side of
these OPEs are given by the Euler beta function of arbitrary weights
for the above operators.
By analyzing the singular behavior of this function
with an appropriate limiting procedure,
the OPEs between the conformally soft positive-helicity
gluon and graviton operators
are obtained and the three corresponding commutator relations
are determined \cite{GHPS}.
The structure constants in these commutators
are quite complex functions of the operator weights and modes.
These structures are simplified through absorption of the various gamma functions (which depend on
the weights and modes) into each current
\cite{Strominger}. The commutator between the gravitons
can be interpreted as the wedge subalgebra of $w_{1+\infty}$ algebra
\cite{Bakas}. The wedge means that
the mode can vary between
one minus the weight and the weight minus one.
The mode of the graviton contains both half integers and integers
while the graviton considered in \cite{Bakas} is an arbitrary
integer. 
The operators in the commutators \cite{GHPS,Strominger}
are dependent on the complex coordinate $z$ of the celestial sphere.
In \cite{HPS}, the mode expansion for the graviton
in the holomorphic sector is performed further
and this leads to an additional contour integral during the
calculation of the commutator relation. Consequently, we obtain a commutator that is independent of the above $z$ coordinate.
Additional details are provided in \cite{Jiang2108,Jiang2110,AMS}
and review papers in \cite{Raclariu,Pasterski,Aneeshetal}
on the celestial holography \footnote{See also Strominger's talk in
strings 2021.}.

The $w_{\infty}$ algebra \cite{Bakas},
as an extension of the Virasoro algebra,
is the Lie algebra where the currents have the conformal weight (or spin)
$s=2,3,4, \cdots, \infty$ \footnote{
This algebra admits the usual central extension in the Virasoro sector
\cite{PRS1990} and can also be obtained
from the contraction \cite{Pope1991}
of the
$W_{\infty}$ algebra \cite{PRS1990,PRS1990-1}, which admits
the central extensions for all sectors of arbitrary conformal
weights.}.
By introducing the conformal weight $s=1$ further into the
$W_{\infty}$ algebra, the $W_{1+\infty}$ algebra is found in \cite{PRS1990-2}
and the complete expression is also given by \cite{Pope1991}.
The structure constants in
the
$W_{\infty}$ algebra are different from those in
the $W_{1+\infty}$ algebra. 
After taking the zero limit of a parameter, we obtain
the Lie algebra between the currents
of weights $s=1,2, \cdots, \infty $
having the central extension in the Virasoro sector.
This algebra will be denoted as $w_{1+\infty}$ algebra.
See also \cite{CL}.
In this paper, we consider the 
$w_{1+\infty}$ algebra with a vanishing central term.

In \cite{OS}, the additional adjoint currents of weights
$s=1,2, \cdots, \infty $ under the $SU(N)$ are added to the above
$W_{1+\infty}$ algebra\footnote{ An important requirement is that
the weight $1$ adjoint current
should produce the affine $\hat{SU}(N)$ algebra with a level $k$
and the adjoint currents should transform under the zero mode
of the weight $1$ adjoint current. After satisfying this requirement and
using the Jacobi identities between the currents,
two additional commutator relations in addition to the
known commutator relation in the $W_{1+\infty}$ algebra
are completely fixed.
The central charge of the Virasoro current is given by
$c = N k$. In addition, the commutator between the adjoint currents
consists of the symmetric tensor $d$ symbol-dependent terms
and the (antisymmetric) structure-constant dependent terms.}.
We expect that
the $w_{1+\infty}$ algebra with $SU(N)$ symmetry will be obtained after
taking the proper limit of a parameter \cite{Sezgin1992}.
In general, a central term from the commutator
between the weight $1$ adjoint currents and
a central term of the Virasoro current are present.

The ${\cal N}=2$ supersymmetric extension of $w_{\infty}$ algebra
is studied in \cite{PS1990} by generalizing the ${\cal N}=2$
superconformal algebra and can be expressed in terms of graded
Poisson brackets along the line of \cite{Bakas}.
No central term is generated.
However, the ${\cal N}=2$ supersymmetric extension of $W_{\infty}$ algebra
\cite{BPRSS}, where the bosonic sector is given by the
sum of $W_{\infty}$ algebra and $W_{1+\infty}$ algebra, exists \footnote{
By taking the proper limit of the parameter, 
the corresponding ${\cal N}=2$
$w_{\infty}$ algebra \cite{Sezgin1991},
where (anti)commutator relations with central terms
comparable to those reported in \cite{PS1990}, is obtained.}.
Based on the construction of the twisted ${\cal N}=2$ superconformal
algebra \cite{Witten,EY}, the topological $W_{\infty}$ algebra \cite{PRSS}
is obtained by twisting
the ${\cal N}=2$ supersymmetric extension of $W_{\infty}$ algebra.
No central term is generated and the structure constants
appearing in two non-trivial commutator relations are the same
\footnote{
Through the contraction procedure (introducing new currents
with a parameter and taking this parameter to be zero),
the right-hand sides of two commutators are simplified
(the anticommutator between the fermionic currents vanishes).
See also \cite{Horava}.}.

In this paper, we generalize the results of \cite{GHPS,Strominger}
to the supersymmetric Einstein-Yang-Mills theory studied in
\cite{FSTZ} where the additional nontrivial OPEs are given by
those between bosonic and fermionic operators.
The work of \cite{Jiang2108}
focuses on similar studies but in a different context:
a supersymmetric extension of \cite{GHPS}.
The OPEs between the fermionic operators are regular.
This is rather unusual because in the conventional conformal
field theory they exhibit non-trivial singular behaviors \footnote{ 
A previous study \cite{Sezgin1989}
has presented a supersymmetric extension
of the $w_{\infty}$ algebra, by generalizing the ${\cal N}=1$ superconformal
algebra. However,
we can reduce the supersymmetry to lower symmetry by using the
twisting procedure employed in \cite{PRSS} where the anticommutator
relation between the fermionic currents vanishes.}.

By using the known results \cite{FSTZ} on the
OPEs between the graviton, gravitino,
gluon, and gluino
associated with the above theory, we would like to describe
the supersymmetric $w_{1+\infty}$ symmetry in terms of the celestial
conformal field theory.

\begin{itemize}
\item[]
The four additional commutator relations
between the soft currents are calculated explicitly
by following the procedures of \cite{GHPS,Strominger} and
focusing on the particular mode of the soft currents (see
for example (\ref{modeH})) along the line of \cite{HPS}:this is a different
viewpoint from \cite{Jiang2108}.
One of the commutators
having no $SU(N)$ group indices can be extracted from
the supersymmetric topological $w_{\infty}$ algebra \cite{PRSS}.
The remaining three can be determined by analyzing
the previous works (\cite{OS} and \cite{PRSS}) further
and introducing the additional symmetry current,
which corresponds to the celestial gluino operator.
The $SU(N)$ symmetry, the supersymmetry and the basic
property of the OPE in the two-dimensional conformal field theory
are used.
Eventually we find that all seven commutator relations
can be identified with those in
the supersymmetric $w_{1+\infty}$ algebra with wedge modes
having the $SU(N)$ symmetry.
\end{itemize}

In section $2$,
we calculate four commutators for the soft currents.
In section $3$,
we present the supersymmetric $w_{1+\infty}$ algebra
corresponding to seven commutators from the soft currents.
In section $4$,
we summarize the main result of this study and discuss ideas for
future work.
In Appendices $A$ and $B$, we repeat the result of \cite{GHPS}
and four commutator relations are successively described.

Various works
\cite{Ahn2107,Ahn2011,AK2009,AKK1910,EP,CHU,CH1906,CH1812}
have been reviewed.
As we will see in section $3$, the work of \cite{OS} and
${\cal N}=2$
$W_{\infty}$ algebra \cite{BPRSS} are useful for understanding
the structure characterizing  the supersymmetric extension of $w_{1+\infty}$
algebra with $SU(N)$. These works are related to
the extension of \cite{OS} or
the ${\cal N}=4$ supersymmetric extension of \cite{BPRSS}.
However, the observation of these algebra
in the context of the celestial conformal field theory remains unexplored.

What we have done or added
in this paper, compared to the previous works
in \cite{HPS,Jiang2108}, is as follows.
In \cite{HPS}, the mode for the graviton current is restricted to
the case in which the corresponding transformations do not mix
$SL(2,R)_L$ primaries and descendants in the OPE between the graviton
currents and matter fields. As described before, these modes are
independent of the complex coordinate $z$ after an integration over
the holomorphic sector further. As noted by \cite{HPS},
we perform the various OPEs between the soft currents and
obtain the commutation relations on these modes.
On the other hand, in \cite{Jiang2108}, as mentioned before,
the supersymmetric generalization of \cite{GHPS} is obtained.
The OPEs between the chiral currents
have simple (or first order) poles in the holomorphic
sector. The modes are labelled by their transformation
under $SL(2,R)_R$ and are dependent of the
complex coordinate $z$.
Each chiral current in \cite{GHPS,Jiang2108} from soft symmetry current
is obtained by taking the multiple derivatives in the
antiholomorphic sector. By construction, these chiral currents
do not depend on the complex coordinate $\bar{z}$.
In this paper, we are focusing on the modes
which do not depend on both $z$ and $\bar{z}$. 
Our aim is to follow the procedure of \cite{GHPS}
and analyze the description of \cite{Strominger}
for the other relevant OPEs
between the soft currents in the context of \cite{HPS}.

\section{ A supersymmetric Einstein-Yang-Mills theory}

We will consider the OPEs in \cite{FSTZ} and obtain the commutator
relations for soft current algebra.

\subsection{A soft current algebra between the graviton and the
gluon: A review}

From the positive-helicity (conformally primary) graviton operator
$G_{\Delta}^+(z,\bar{z})$
with two-dimensional conformal weight $\Delta$,
a family of (conformally) soft positive-helicity graviton current
is defined as \cite{GHPS}
\bea
H^k(z,\bar{z}) = \lim_{\varepsilon \rightarrow 0} \varepsilon \,
G^{+}_{k+\varepsilon}(z,\bar{z}), \qquad
k =2, 1, 0, -1, -2, \cdots,
\label{HandG}
\eea
where, the (celestial) left and right conformal weights are given by
\bea
(h, \bar{h}) = \Big(\frac{k+2}{2}, \frac{k-2}{2}\Big).
\label{handhbar}
\eea
The additional factor of
$\varepsilon$ in (\ref{HandG}) is necessary for canceling the pole
of the beta function appearing in the original OPE coefficient. 

Taking the holomorphic and antiholomorphic
expansions for the above soft graviton current we obtain:
\bea
H^k(z,\bar{z})
=  \sum_{n=\frac{k-2}{2}}^{\frac{2-k}{2}}\,
\frac{H_{n}^k(z)}{\bar{z}^{n+\frac{k-2}{2}}}
= \sum_{m}\, \sum_{n=\frac{k-2}{2}}^{\frac{2-k}{2}}\,
\frac{H_{m,n}^k}{z^{m+\frac{k+2}{2}}\, \bar{z}^{n+\frac{k-2}{2}}}.
\label{H}
\eea
Rather than using the $H_{n}^k(z)$ mode,
which depends on the holomorphic coordinate $z$,
we further expand the current with respect to the holomorphic
mode $m$ in order to obtain the closed algebra with
the $SL(2,R)_R$ generators \cite{HPS}.
The operator $H_{m,n}^k$ is therefore independent of $z$ and $\bar{z}$.
In this work, we focus on the case where the mode $m$ is equal to
$(1-h)$ together with (\ref{handhbar})
\bea
\hat{H}_{n}^k \equiv H_{m=1-h,n}^k.
\label{modeH}
\eea
This will lead to $\frac{1}{z}$ dependence for particular
terms of (\ref{H}).
As usual, the mode in (\ref{modeH})
can be expressed in terms of the following
contour integral
\bea
\hat{H}_{n}^k = \oint_{|z| < \varepsilon} \, \, \frac{d z}{2\pi i}\,
z^{1-\frac{k+2}{2}+\frac{k+2}{2}-1}\,
\oint_{|\bar{z}| < \varepsilon} \,
\frac{d \bar{z}}{2\pi i}\, \bar{z}^{n+\frac{k-2}{2}-1}\,
H^k(z, \bar{z}),
\label{cont}
\eea
where, we intentionally express the power of $z$ explicitly
in the integrand. This can be easily checked
(\ref{cont}) by substituting the relation (\ref{H})
into the right-hand side of (\ref{cont}).

We can repeat the computation
performed in \cite{GHPS}. For the calculation of
$\big[ \hat{H}_m^p, \hat{H}_n^q \big]$, we should perform
the contour integrals over $z_1, \bar{z}_1, z_2$,
and $\bar{z}_2$ with the OPE
$H^{k}(z_1,\bar{z}_1)\, H^l(z_2,\bar{z}_2)$
in addition to some powers of $\bar{z}_1$ and $\bar{z}_2$.
The three contour integrals (except for the coordinate $z_2$)
can be done exactly without any modification. From this procedure,
we are left with the contour integral over $z_2$
acting on the $\sum_{p}\, \frac{H^{k+l}_{p,m+n}}{
z_2^{p+\frac{k+l+2}{2}}}$ as well as the mode- and weight-dependent
terms.
This leads to $\hat{H}_{m+n}^{k+l}$ being generated by the 
$\frac{1}{z_2}$ factor.
We present the explicit commutator relation between the
soft graviton currents in Appendix $A$. Similarly,
the commutator between
the soft gluon currents and the commutator
between the soft graviton current and the soft gluon current
can be determined (see Appendix (\ref{GHPSresult})).

In \cite{Strominger}, 
the absorption of the
mode- and weight-dependent terms appearing in the 
right-hand side denominator
of the commutator relation
are systematically investigated. The corresponding
factors in the numerator
can then be absorbed in the soft current of the right side.
In other words, we have \cite{Strominger,HPS}
\bea
\hat{w}_n^p \equiv \frac{1}{\kappa}\, (p-n-1)! \, (p+n-1)!
\, \hat{H}_n^{-2p+4}, \qquad
\hat{J}_m^{q,a} \equiv (q-m-1)! \, (q+m-1)!\, \hat{R}_m^{3-2q,a}, 
\label{hatone}
\eea
where, $\kappa$ is the gravitational coupling constant
and the index $a$ is an adjoint index of $SU(N)$.

Therefore, with the help of (\ref{hatone}),
we have \footnote{
Equations $(2.7)$, $(3.6)$, and $(3.8)$ reported in \cite{Strominger}
correspond to (\ref{softalgebra}).} 
\bea
\big[ \hat{w}_m^p, \hat{w}_n^q \big] & = &
\Big[ m(q-1)-n(p-1)\Big]\, \hat{w}_{m+n}^{p+q-2}, 
\nonu \\
\big[ \hat{J}_m^{p,a}, \hat{J}_n^{q,b} \big] & = & -i \, f^{a b}_{\,\,\,\,c}
\,  \hat{J}_{m+n}^{p+q-1,c}, 
\nonu \\
\big[ \hat{w}_m^p, \hat{J}_n^{q,a} \big] & = & \Big[ m(q-1)-n(p-1)\Big]\,
\hat{J}_{m+n}^{p+q-2, a}. 
\label{softalgebra}
\eea
As observed in \cite{Strominger,HPS}, 
each $\hat{w}^q$
is associated with a finite number of
modes $1-q \leq n \leq q-1$, which provide $(2 q-1)$ dimensional closed
algebra, and $\hat{w}^{q=2}$
serves as a $SL(2,R)_R$ generator.
Here, $q$ is the positive half integer value
$q=1,\frac{3}{2}, 2, \frac{5}{2}, \cdots $.
We fix $p=2$ in the first equation of (\ref{softalgebra}),
and this implies that
the $n$-th mode of a weight $q$  transforms 
as a primary under the $SL(2,R)_R$ generator $\hat{w}^2_m$.
From the third equation of (\ref{softalgebra}),
we check that 
the $n$-th mode of a weight $q$  transforms 
as a primary under the $\hat{w}^2_m$ and
$q$ runs over 
$q=1,\frac{3}{2}, 2, \frac{5}{2}, \cdots $ as previously mentioned.
The first equation of (\ref{softalgebra}) includes
the wedge subalgebra of $w_{1+\infty}$ algebra \cite{Bakas}.
We will provide the corresponding description in the conventional conformal
field theory outlined in the subsequent section.

\subsection{Further soft current algebra in the presence of
gravitino and gluino currents }

We continue our calculation of the soft current algebra
and account for the occurrence of fermionic currents.

\subsubsection{The commutator between the graviton and the
gravitino}

From the positive-helicity (conformally primary) gravitino operator
${\cal O}_{\Delta,+\frac{3}{2}}(z,\bar{z})$
with two-dimensional conformal weight $\Delta$,
a family of (conformally) soft positive-helicity gravitino current
is defined as \cite{Jiang2108}
\bea
I^k(z,\bar{z}) = \lim_{\varepsilon \rightarrow 0} \varepsilon \,
{\cal O}_{k+\varepsilon,+\frac{3}{2}}(z,\bar{z}), \qquad
k =\frac{3}{2}, \frac{1}{2}, -\frac{1}{2}, -\frac{3}{2}, \cdots,
\label{IandO}
\eea
where the (celestial) left and right conformal weights are given by
\footnote{By performing the supersymmetric Ward identities \cite{FSTZ}
successively, the sub-subleading graviton $(\Delta=-1)$
leads to the subleading gravitino $(\Delta=-\frac{1}{2})$.
Now the latter provides the subleading graviton $(\Delta=0)$
which enables us to obtain the leading gravitino
$(\Delta=\frac{1}{2})$. Finally, we arrive at the leading graviton
$(\Delta =1)$ from the latter.
Then we should exclude $k=2$ for graviton and $k=\frac{3}{2}$
for gravitino as soft currents \cite{Jiang2108}.
}
\bea
(h, \bar{h}) = \Big(\frac{k+\frac{3}{2}}{2}, \frac{k-\frac{3}{2}}{2}\Big).
\label{handhbar1}
\eea
The OPE of
the conformal primary graviton and
the conformal primary gravitino
of arbitrary weights is given as follows \cite{FSTZ}:
\bea
&& {\cal O}_{\Delta_1, +2}(z_1, \bar{z}_1)\,
{\cal O}_{\Delta_2, +\frac{3}{2}}(z_2,\bar{z}_2)=
\nonu \\
&&
-\frac{\frac{\kappa}{2}}{z_{12}}\, \sum_{n=0}^{\infty}\,
B(\Delta_1-1+n, \Delta_2-\frac{1}{2})\,
\frac{\bar{z}_{12}^{n+1}}{n!}\, \bar{\pa}^n \, 
{\cal O}_{\Delta_1+\Delta_2, +\frac{3}{2}}(z_2,\bar{z}_2) + \cdots.
\label{OOope1}
\eea
The abbreviated parts contain the left-conformal descendants.
In the corresponding expression of \cite{FSTZ},
the $\bar{z}_{12}$ in the numerator is moved into
the inside of the summation in (\ref{OOope1}).
We express the OPE with the bosonic operators
located at a position of $(z_1, \bar{z}_1)$, rather than
at $(z_2,\bar{z}_2)$, in order to use the previous relations
on the finite sum.

The OPE between the soft positive-helicity graviton (\ref{HandG}) and the
soft positive-helicity gravitino (\ref{IandO})
can be described by 
\bea
H^k(z_1,\bar{z}_1)\, I^l(z_2,\bar{z}_2)=
-\frac{\kappa}{2}\, \frac{1}{z_{12}}\,
\sum_{n=0}^{1-k}\,
\left(
\begin{array}{c}
1-n + \frac{1}{2}-k-l \\
\frac{1}{2}-l
\end{array}
\right)\,
\frac{\bar{z}_{12}^{n+1}}{n!}\, \bar{\pa}^n \, 
I^{k+l}(z_2,\bar{z}_2) + \cdots.
\label{HIope}
\eea
The bracket denotes a binomial coefficient
\footnote{ Note that the right-hand side of (\ref{HIope})
looks very similar to equation $(3.5)$ of \cite{GHPS}
in the sense that after we replace $l$ with $l+\frac{1}{2}$
we obtain (\ref{HIope}).}.
Consider the maximum value for the dummy variable $n$
in the summation of (\ref{HIope}) from the infinite sum
in (\ref{OOope1}). This value stems from the fact that
the highest power of ${\bar{z}_1}$ in (\ref{H}) is $(2-k)$.
We then obtain the relation $\pa^{(3-k)}_{\bar{z}_1} \, H^k(z_1,\bar{z}_1)
=0$ and in the right-hand side of (\ref{HIope}),
the highest power of $\bar{z}_{12}$ should also be $(2-k)$.

The corresponding commutator relation
is given as \footnote{
The mode expansion is given as follows:
$I^l(z,\bar{z})
=  \sum_{n=\frac{l-\frac{3}{2}}{2}}^{\frac{\frac{3}{2}-l}{2}}\,
\frac{I_{n}^l(z)}{\bar{z}^{n+\frac{l-\frac{3}{2}}{2}}}
= \sum_{m}\, \sum_{n=\frac{l-\frac{3}{2}}{2}}^{\frac{\frac{3}{2}-l}{2}}\,
\frac{I_{m,n}^l}{z^{m+\frac{l+\frac{3}{2}}{2}}\, \bar{z}^{n+\frac{l-\frac{3}{2}}{2}}}
$ with (\ref{handhbar1})
and  $\hat{I}_{n}^{l} \equiv I_{1-\frac{l+\frac{3}{2}}{2},n}^{l}$.}
\bea
\big[ \hat{H}_m^k, \hat{I}_n^l \big] & = &
\oint_{|\bar{z}_1| < \varepsilon} \, \frac{d \bar{z}_1}{2\pi i}\,
\bar{z}_1^{m+\frac{k-2}{2}-1}\,
\oint_{|\bar{z}_2| < \varepsilon} \, \frac{d \bar{z}_2}{2\pi i}\,
\bar{z}_2^{n+\frac{l-\frac{3}{2}}{2}-1}\,
\oint_{|z_{12} | < \varepsilon} \,  \frac{d z_1}{2\pi i}\,
\oint_{|z_2| < \varepsilon} \,  \frac{d z_2}{2\pi i}\,
\nonu \\
&\times & H^k(z_1,\bar{z}_1)\, I^l(z_2,\bar{z}_2).
\label{HIcomm}
\eea
The
$\bar{z}_1$ and $z_1$ contours receive unequal treatment due to the singular term of
$\frac{1}{z_{12}}$ in (\ref{HIope}).
This can be compared with the approach of \cite{HPS}
that allows equal treatment of these contours.
After inserting the relation (\ref{HIope}) into the relation
(\ref{HIcomm}), four contour integrals are generated.
The contour integral over the $z_1$ coordinate with a factor
$\frac{1}{2 \pi i}$ where the corresponding integrand is
$\frac{1}{z_{12}}$ is simply one \footnote{
The contour integral over the $\bar{z}_1$ coordinate can be
obtained by using the Appendix $(A.7)$ identity of \cite{GHPS}.
Moreover, the  subsequent contour integral over the $\bar{z}_{2}$ coordinate
can be obtained by using the identity presented in Appendix $(A.9)$
of \cite{GHPS}.}.
The contour integral over the coordinate is then given as:
$z_2$ selects $\hat{I}_{m+n}^{k+l}$.
From this we obtain: 
\bea
\big[ \hat{H}_m^k, \hat{I}_n^l \big] & = &
-\frac{\kappa}{2}\,
\frac{(-1)^{m+\frac{k}{2}}\, (-m-n-\frac{k+l-\frac{3}{2}}{2})!}{
(\frac{1}{2}-l)!\, (\frac{2-k}{2}-m)!} \nonu \\
& \times & 
\sum_{s=-m-\frac{k}{2}}^{1-k}\, \frac{(-1)^s \, (s+1)\,
(\frac{3}{2}-s-k-l)!}{(1-s-k)!\, (s+m+\frac{k}{2})!\, (-m-n-
\frac{k+l-\frac{3}{2}}{2}-s)!}\,
\hat{I}_{m+n}^l.
\label{commHI}
\eea
As we expected, 
this intermediate result is the same as
that presented in \cite{GHPS} when we replace $l$ with $(l+\frac{1}{2})$.
See also the first identity appearing in the footnote \ref{iden}.
Currently, obtaining a closed-form expression
of the finite sum over the dummy variable
$s$ is challenging.
however, we can take the
expression obtained from \cite{GHPS} (or previously mentioned first identity)
and substitute
several values for the modes and weights into the relation.
This allows expression of the above finite sum in terms of gamma functions.

The above result (\ref{commHI}) can be reduced to
\bea
\big[\hat{H}_m^{k}, \hat{I}_n^{l} \big] & = & \frac{\kappa}{2}\,
\Big[ m(\frac{3}{2}-l)-n(2-k)\Big] \nonu \\
& \times &  
\frac{(\frac{2-k}{2} -m +\frac{\frac{3}{2}-l}{2} -n-1 )!\,
(\frac{2-k}{2} +m +\frac{\frac{3}{2}-l}{2} +n-1 )!}
{(\frac{2-k}{2} -m)!\,
(\frac{\frac{3}{2}-l}{2} -n)! \,
(\frac{2-k}{2} +m)! \, (\frac{\frac{3}{2}-l}{2} +n)!}
\, \hat{I}_{m+n}^{k+l}.
\label{hicomm}
\eea
We observe that
the numerical
mode- and weight-dependent factor in the right-hand side of
(\ref{hicomm}) 
is the same as the factor included in
$\big[\hat{H}_m^{k}, \hat{H}_n^{l} \big]$
where the weight $l$ is replaced by $(l+\frac{1}{2})$.
Further details are provided in Appendix (\ref{theresult}).

As done in (\ref{hatone}), we would like to
absorb the denominator of (\ref{hicomm}) into the
currents by changing the weights with mode-dependent parts.
Together with the first relation of (\ref{hatone}),
we introduce the following similar quantity
\bea
\hat{G}_n^{q} \equiv \frac{1}{\kappa}\,
(q-n-1)! \, (q+n-1)!\, \hat{I}_n^{\frac{7}{2}-2q}. 
\label{hatone1}
\eea
The above commutator can then be summarized as follows:
\bea
\big[ \hat{w}_m^{p}, \hat{G}_n^{q} \big] & = &
\Big[ m(q-1)-n(p-1)\Big]\,
\hat{G}_{m+n}^{p+q-2}.
\label{fincomm}
\eea
The $n$-th mode of a weight $q$ in (\ref{fincomm}) transforms 
as a primary under the $\hat{w}^2_m$. Furthermore, $q$ runs over 
$q=1, \frac{3}{2}, 2, \frac{5}{2}, \cdots $ and its mode $n$ varies
as $1-q\leq n \leq q-1$.
The $m,n,p$ and $q$ dependence observed here
is the same as that included in (\ref{softalgebra}).

\subsubsection{The commutator between the gluon and the
gluino}

From the positive-helicity (conformally primary) gluino operator
${\cal O}_{\Delta,+\frac{1}{2}}(z,\bar{z})$
with two-dimensional conformal weight $\Delta$,
a family of (conformally) soft positive-helicity gluino current
is defined as \cite{Jiang2108}:
\bea
L^{k,a}(z,\bar{z}) = \lim_{\varepsilon \rightarrow 0} \varepsilon \,
{\cal O}^a_{k+\varepsilon,+\frac{1}{2}}(z,\bar{z}), \qquad
k =\frac{1}{2}, -\frac{1}{2}, -\frac{3}{2}, \cdots,
\label{LandO}
\eea
where, the (celestial) left and right conformal weights are given by
\bea
(h, \bar{h}) = \Big(\frac{k+\frac{1}{2}}{2}, \frac{k-\frac{1}{2}}{2}\Big).
\label{handhbar2}
\eea
The OPE of
the (conformal primary) gluon and
the (conformal primary) gluino
of arbitrary weights is given as follows \cite{FSTZ}:
\bea
&& {\cal O}^a_{\Delta_1, +1}(z_1, \bar{z}_1)\,
{\cal O}^b_{\Delta_2, +\frac{1}{2}}(z_2,\bar{z}_2)=
\nonu \\
&& \frac{-i \, f^{a b}_{\,\,\,\,c}}{z_{12}}\, \sum_{n=0}^{\infty}\,
B(\Delta_1-1+n, \Delta_2-\frac{1}{2})\,
\frac{\bar{z}_{12}^{n}}{n!}\, \bar{\pa}^n \, 
{\cal O}^c_{\Delta_1+\Delta_2-1, +\frac{1}{2}}(z_2,\bar{z}_2) + \cdots.
\label{OOope2}
\eea
The regular term $\de^{a b}$ reported in \cite{FSTZ} is neglected here. 

The OPE between the soft positive-helicity gluon and the
soft positive-helicity gluino (\ref{LandO})
can be summarized as follows:
\bea
R^{k,a}(z_1,\bar{z}_1)\, L^{l,b}(z_2,\bar{z}_2)=
\frac{- i \, f^{a b }_{\,\,\,\,c}}{z_{12}}\,
\sum_{n=0}^{1-k}\,
\left(
\begin{array}{c}
1-n + \frac{1}{2}-k-l \\
\frac{1}{2}-l
\end{array}
\right)\,
\frac{\bar{z}_{12}^{n}}{n!}\, \bar{\pa}^n \, 
L^{k+l-1,c}(z_2,\bar{z}_2) + \cdots.
\label{RLope}
\eea
The structure constant $f^{a b}_{\,\,\,\,c}$
is associated with the $SU(N)$.
Note that the infinite sum presented in (\ref{OOope2})
is reduced to the finite sum due to the fact that
$\pa^{(2-k)}_{\bar{z}_1} \, R^{k,a}(z_1,\bar{z}_1)
=0$
\footnote{ As observed in a previous subsection,
this OPE (\ref{RLope}) looks very similar to equation $(2.7)$
of \cite{GHPS} and the binomial coefficient where $l$ is replaced by
$(l+\frac{1}{2})$ becomes the above expression.}. 
The corresponding commutator relation
\footnote{
The mode expansion is given by
$L^{l,b}(z,\bar{z})
=  \sum_{n=\frac{l-\frac{1}{2}}{2}}^{\frac{\frac{1}{2}-l}{2}}\,
\frac{L_{n}^{l,b}(z)}{\bar{z}^{n+\frac{l-\frac{1}{2}}{2}}}
= \sum_{m}\, \sum_{n=\frac{l-\frac{1}{2}}{2}}^{\frac{\frac{1}{2}-l}{2}}\,
\frac{L_{m,n}^{l,b}}{z^{m+\frac{l+\frac{1}{2}}{2}}\, \bar{z}^{n+\frac{l-\frac{1}{2}}{2}}}
$ with (\ref{handhbar2})
and $\hat{L}_{n}^{l,b} \equiv L_{1-\frac{l+\frac{1}{2}}{2},n}^{l,b}$.}
can then be written in terms of
\bea
\big[ \hat{R}_m^{k,a}, \hat{L}_n^{l,b} \big] & = &
\oint_{|\bar{z}_1| < \varepsilon}  \, \frac{d \bar{z}_1}{2\pi i}\,
\bar{z}_1^{m+\frac{k-1}{2}-1}\,
\oint_{|\bar{z}_2| < \varepsilon}  \, \frac{d \bar{z}_2}{2\pi i}\,
\bar{z}_2^{n+\frac{l-\frac{1}{2}}{2}-1}\,
\oint_{|z_{12}| < \varepsilon}  \,  \frac{d z_1}{2\pi i}\,
\oint_{|z_2| < \varepsilon}  \,  \frac{d z_2}{2\pi i}\,
\nonu \\
& \times & 
R^{k,a}(z_1,\bar{z}_1)\, L^{l,b}(z_2,\bar{z}_2).
\label{RLcomm}
\eea

After we calculate the contour integrals over
$z_1$, $\bar{z}_1$,$\bar{z}_2$, and $z_2$ successively,
we obtain the following
intermediate result from (\ref{RLcomm})
\bea
\big[ \hat{R}_m^{k,a}, \hat{L}_n^{l,b} \big] & = &
-i \, f^{a b}_{\,\,\,\,c}\,
\frac{(-1)^{m+\frac{k-1}{2}}\,
(\frac{1-k}{2}-m+\frac{\frac{1}{2}-l}{2}-n)!}{
(\frac{1}{2}-l)!\, (\frac{1-k}{2}-m)!}
\label{commRL} \\
& \times & 
\sum_{s=-m+\frac{1-k}{2}}^{1-k}\, \frac{(-1)^s \,
(\frac{3}{2}-s-k-l)!}{(1-s-k)!\, (s+m+\frac{k-1}{2})!\,
(\frac{1-k}{2}-m+\frac{\frac{1}{2}-l}{2}-n-s)!}\,
\hat{L}_{m+n}^{k+l-1,c}.
\nonu
\eea
See also the second identity appearing in the footnote \ref{iden}
\footnote{As mentioned before, the above finite sum in (\ref{commRL})
can be read off
from the analysis of the equation presented in Appendix $(A.8)$
of
\cite{GHPS} where the replacement of $l$ and $(l+\frac{1}{2})$ is addressed.}. 
Although observation of the closed form
in terms of gamma functions is challenging, 
we can check the corresponding identity by
applying several values for the modes and weights.
This yields: 
\bea
\big[\hat{R}_m^{k,a}, \hat{L}_n^{l,b} \big] & = & - i \, f^{ab}_{\,\,\,\,\,c}\,
\frac{(\frac{1-k}{2} -m +\frac{\frac{1}{2}-l}{2} -n )!\,
(\frac{1-k}{2} +m +\frac{\frac{1}{2}-l}{2} +n )!}{(\frac{1-k}{2} -m)!\,
(\frac{\frac{1}{2}-l}{2} -n)! \, (\frac{1-k}{2} +m)! \,
(\frac{\frac{1}{2}-l}{2} +n)!}
\, \hat{L}_{m+n}^{k+l-1,c}.
\label{rlcomm}
\eea

In order to simplify the commutation relation of (\ref{rlcomm}),
we introduce
\bea
\hat{\psi}_n^{q,b} \equiv (q-n-1)! \, (q+n-1)!\,
\hat{L}_n^{\frac{5}{2}-2q,b}, 
\label{hatone2}
\eea
together with the second relation of (\ref{hatone}).
Afterward, we obtain the final commutator relation as follows:
\bea
\big[ \hat{J}_m^{p,a}, \hat{\psi}_n^{q,b} \big] & = &
- i \, f^{a b }_{\,\,\,\,c}\,
\hat{\psi}_{m+n}^{p+q-1,c}.
\label{fin1comm}
\eea
We will
observe, in the subsequent subsection, that
the $n$-th mode of a weight $q$  in (\ref{fin1comm}) transforms 
as a primary under the $\hat{w}^2_m$. In addition,
$q$ runs over 
$q=1, \frac{3}{2}, 2, \frac{5}{2}, \cdots $ and
its mode $n$ varies
as $1-q\leq n \leq q-1$, as in a previous case. Note that
the weight of the right-hand side of (\ref{fin1comm}) is given by $(p+q-1)$
as in the second case of (\ref{softalgebra}).

\subsubsection{The commutator between the gluon and the
gravitino}

The OPE of
the (conformal primary) gluon and
the (conformal primary) gravitino
of arbitrary weights is given as follows \cite{FSTZ}:
\bea
&& {\cal O}^a_{\Delta_1, +1}(z_1, \bar{z}_1)\,
{\cal O}_{\Delta_2, +\frac{3}{2}}(z_2,\bar{z}_2)=
\nonu \\
&& -\frac{\kappa}{2}\, \frac{\bar{z}_{12}}{z_{12}}\, \sum_{n=0}^{\infty}\,
B(\Delta_1+n, \Delta_2-\frac{1}{2})\,
\frac{\bar{z}_{12}^{n}}{n!}\, \bar{\pa}^n \, 
{\cal O}^a_{\Delta_1+\Delta_2, +\frac{1}{2}}(z_2,\bar{z}_2)+ \cdots.
\label{OOope3}
\eea
Note that a gluino occurs in the right-hand side of this OPE.
The OPE between the soft positive-helicity gluon and the
soft positive-helicity gravitino
can be obtained from:
\bea
R^{k,a}(z_1,\bar{z}_1)\, I^{l}(z_2,\bar{z}_2)=
-\frac{\frac{\kappa}{2}}{z_{12}}\,
\sum_{n=0}^{1-k}\,
\left(
\begin{array}{c}
-n + \frac{1}{2}-k-l \\
\frac{1}{2}-l
\end{array}
\right)\,
\frac{\bar{z}_{12}^{n+1}}{n!}\, \bar{\pa}^n \, 
L^{k+l,a}(z_2,\bar{z}_2) + \cdots.
\label{RIope}
\eea
The corresponding commutator relation can be obtained from
(\ref{RIope}) with various contour integrals.
In the expression reported in (\ref{OOope3}), the additional factor
$\bar{z}_{12}$ is joined in the inside of the summation
presented in (\ref{RIope}).
As in a previous case, by determining the power of $\bar{z}_1$ and $\bar{z}_2$
in the integrand via the conformal weights of
currents, we obtain the following expression: 
\bea
\big[ \hat{R}_m^{k,a}, \hat{I}_n^{l} \big] & = &
\oint_{|\bar{z}_1| < \varepsilon}  \, \frac{d \bar{z}_1}{2\pi i}\,
\bar{z}_1^{m+\frac{k-1}{2}-1}\,
\oint_{|\bar{z}_2| < \varepsilon}  \, \frac{d \bar{z}_2}{2\pi i}\,
\bar{z}_2^{n+\frac{l-\frac{3}{2}}{2}-1}\,
\oint_{|z_{12}| < \varepsilon}  \,  \frac{d z_1}{2\pi i}\,
\oint_{|z_2| < \varepsilon}  \,  \frac{d z_2}{2\pi i}\,
\nonu \\
& \times & R^{k,a}(z_1,\bar{z}_1)\, I^{l}(z_2,\bar{z}_2).
\label{RIcomm}
\eea

By substituting the OPE in (\ref{RIope}) into
(\ref{RIcomm}) and performing each contour integral
successively, we obtain the following intermediate result,
which consists of the finite sum with
the mode and weight-dependent overall factor, 
\bea
\big[ \hat{R}_m^{k,a}, \hat{I}_n^{l} \big] & = &
-\frac{\kappa}{2}\,
\frac{(-1)^{1+m+\frac{k-1}{2}}\,
(-m-n-\frac{k+l-\frac{1}{2}}{2})!}{
(\frac{1}{2}-l)!\, (\frac{1-k}{2}-m)!}
\label{commRLI} \\
& \times & 
\sum_{s=-1-m+\frac{1-k}{2}}^{1-k}\, \frac{(-1)^s \,
(s+1)\, (\frac{1}{2}-s-k-l)!}{(-s-k)!\, (1+s+m+\frac{k-1}{2})!\,
(-m-n-\frac{k+l-\frac{1}{2}}{2}-s)!}\,
\hat{L}_{m+n}^{k+l,a}.
\nonu
\eea
Note that the above finite sum with $l$ replaced by
$(l-\frac{1}{2})$ in (\ref{commRLI}) is calculated in \cite{GHPS}
to determine the commutator between the soft gluon and soft graviton.
\footnote{
The corresponding identity reported in \cite{GHPS} is given as:
$\sum_{s=-1-m-\frac{k-1}{2}}^{1-k}\, \frac{(-1)^s \, (s+1)\,
(1-s-k-l)!}{(1-s-k)!\, (s+m+\frac{k}{2})!\, (-m-n-
\frac{k+l-2}{2}-s)!} =
\Big[ -m(2-l)+n(1-k)\Big]
\,
\frac{(1-l)!\,
(\frac{1-k}{2} +m +\frac{2-l}{2} +n-1 )!}
{(-1)^{1+m+\frac{k-1}{2}}\,
(\frac{2-l}{2} -n)! \,
(\frac{1-k}{2} +m)! \, (\frac{2-l}{2} +n)!}$.}.
See also the first identity appearing in the footnote \ref{iden}.
We obtain the final result by using the explicit
form, which includes various gamma functions in the fractional form,
\bea
\big[\hat{R}_m^{k,a}, \hat{I}_n^{l} \big] & = & 
\frac{\kappa}{2}\,
\Big[ m(\frac{3}{2}-l)-n(1-k)\Big] \nonu \\
&\times &  
\frac{(\frac{1-k}{2} -m +\frac{\frac{3}{2}-l}{2} -n-1 )!\,
(\frac{1-k}{2} +m +\frac{\frac{3}{2}-l}{2} +n-1 )!}{(\frac{1-k}{2} -m)!\,
(\frac{\frac{3}{2}-l}{2} -n)! \, (\frac{1-k}{2} +m)! \,
(\frac{\frac{3}{2}-l}{2} +n)!}
\, \hat{L}_{m+n}^{k+l,a}.
\label{ricomm}
\eea

From equations, (\ref{hatone}), (\ref{hatone1}), and (\ref{hatone2}),
the above commutation relation (\ref{ricomm})
is:
\bea
\big[ \hat{J}_m^{p,a}, \hat{G}_n^{q} \big] & = &
\Big[ m(q-1)-n(p-1)\Big]\,
\hat{\psi}_{m+n}^{p+q-2,a}.
\label{fin2comm}
\eea
The $m,n,p$ and $q$ dependence here
is the same as  that occurring in (\ref{softalgebra}).

\subsubsection{The commutator between the graviton and the
gluino}

The OPE of
the (conformal primary) graviton and
the (conformal primary) gluino
of arbitrary weights is given by \cite{FSTZ}
\bea
&& {\cal O}_{\Delta_1, +2}(z_1, \bar{z}_1)\,
{\cal O}^a_{\Delta_2, +\frac{1}{2}}(z_2,\bar{z}_2)=
\nonu \\
&& - \frac{\kappa}{2} \, \frac{\bar{z}_{12}}{z_{12}}\, \sum_{n=0}^{\infty}\,
B(\Delta_1-1+n, \Delta_2+\frac{1}{2})\,
\frac{\bar{z}_{12}^{n}}{n!}\, \bar{\pa}^n \, 
{\cal O}^a_{\Delta_1+\Delta_2, +\frac{1}{2}}(z_2,\bar{z}_2)+\cdots.
\label{OOope4}
\eea
The OPE between the soft positive-helicity graviton and the
soft positive-helicity gluino
can be obtained from:
\bea
H^{k}(z_1,\bar{z}_1)\, L^{l,a}(z_2,\bar{z}_2)=
-\frac{\kappa}{2}\, \frac{1}{z_{12}}\,
\sum_{n=0}^{1-k}\,
\left(
\begin{array}{c}
1-n - \frac{1}{2}-k-l \\
-\frac{1}{2}-l
\end{array}
\right)\,
\frac{\bar{z}_{12}^{n+1}}{n!}\, \bar{\pa}^n \, 
L^{k+l,a}(z_2,\bar{z}_2) + \cdots.
\label{HLope}
\eea
The finite terms from (\ref{OOope4}) survive in the conformally soft
limit
\footnote{From equation $(4.2)$ of \cite{GHPS},
by replacing the $l$ with $(l+\frac{1}{2})$, 
the binomial coefficient becomes the abovementioned value reported in (\ref{HLope}).}.
Again the commutator between the two currents
can be determined as follows:
\bea
\big[ \hat{H}_m^{k}, \hat{L}_n^{l,a} \big] & = &
\oint_{|\bar{z}_1| < \varepsilon}  \, \frac{d \bar{z}_1}{2\pi i}\,
\bar{z}_1^{m+\frac{k-2}{2}-1}\,
\oint_{|\bar{z}_2| < \varepsilon}  \, \frac{d \bar{z}_2}{2\pi i}\,
\bar{z}_2^{n+\frac{l-\frac{1}{2}}{2}-1}\,
\oint_{|z_{12}| < \varepsilon}  \,  \frac{d z_1}{2\pi i}\,
\oint_{|z_2| < \varepsilon}  \,  \frac{d z_2}{2\pi i}\,
\nonu \\
& \times &
H^{k}(z_1,\bar{z}_1)\, L^{l,a}(z_2,\bar{z}_2).
\label{HLcomm}
\eea

We arrive at the intermediate result for the commutator,
from (\ref{HLope}) and (\ref{HLcomm}),
\bea
\big[ \hat{H}_m^{k}, \hat{L}_n^{l,a} \big] & = &
-\frac{\kappa}{2}\,
\frac{(-1)^{m+\frac{k}{2}}\,
(-m-n-\frac{k+l-\frac{1}{2}}{2})!}{
(-\frac{1}{2}-l)!\, (\frac{2-k}{2}-m)!}
\label{commHL} \\
& \times & 
\sum_{s=-m-\frac{k}{2}}^{1-k}\, \frac{(-1)^s \,
(s+1)\, (\frac{1}{2}-s-k-l)!}{(1-s-k)!\, (1+s+m+\frac{k-2}{2})!\,
(-m-n-\frac{k+l-\frac{1}{2}}{2}-s)!}\,
\hat{L}_{m+n}^{k+l,a}.
\nonu
\eea
We realize that the finite sum in (\ref{commHL})
appears in (\ref{commHI}) and by replacing $l$ with
$(l+1)$ the latter becomes the former.
See also the first identity of the footnote \ref{iden}.
Therefore, we obtain:
\bea
\big[\hat{H}_m^{k}, \hat{L}_n^{l,a} \big] & = & \frac{\kappa}{2}\,
\Big[ m(\frac{1}{2}-l)-n(2-k)\Big] \nonu \\
&\times &  
\frac{(\frac{2-k}{2} -m +\frac{\frac{1}{2}-l}{2} -n-1 )!\,
(\frac{2-k}{2} +m +\frac{\frac{1}{2}-l}{2}
+n-1 )!}{(\frac{2-k}{2} -m)!\,
(\frac{\frac{1}{2}-l}{2} -n)! \,
(\frac{2-k}{2} +m)! \, (\frac{\frac{1}{2}-l}{2} +n)!}
\, \hat{L}_{m+n}^{k+l,a}.
\label{hlcomm}
\eea

By using the equations, (\ref{hatone}) and (\ref{hatone2}),
the above commutation relation (\ref{hlcomm}) becomes
\bea
\big[ \hat{w}_m^{p}, \hat{\psi}_n^{q,a} \big] & = &
\Big[ m(q-1)-n(p-1)\Big]\,
\hat{\psi}_{m+n}^{p+q-2,a}. 
\label{fin3comm}
\eea
The $n$-th mode of a weight $q$  in (\ref{fin3comm}) transforms 
as a primary under the $\hat{w}^2_m$. In addition,
$q$ runs over 
$q=1, \frac{3}{2}, 2, \frac{5}{2}, \cdots $ and
its mode $n$ varies
as $1-q\leq n \leq q-1$ as previously stated.

\subsubsection{Summary of this subsection}

We collect the previous four commutator relations,
(\ref{fincomm}), (\ref{fin1comm}), (\ref{fin2comm}),
and (\ref{fin3comm}) as follows:
\bea
\big[ \hat{J}_m^{p,a}, \hat{\psi}_n^{q,b} \big] & = &
- i \, f^{a b }_{\,\,\,\,c}\,
\hat{\psi}_{m+n}^{p+q-1,c}, 
\nonu \\
\big[ \hat{w}_m^{p}, \hat{G}_n^{q} \big] & = &
\Big[ m(q-1)-n(p-1)\Big]\,
\hat{G}_{m+n}^{p+q-2}, 
\nonu \\
\big[ \hat{J}_m^{p,a}, \hat{G}_n^{q} \big] & = &
\Big[ m(q-1)-n(p-1)\Big]\,
\hat{\psi}_{m+n}^{p+q-2,a}, 
\nonu \\
\big[ \hat{w}_m^{p}, \hat{\psi}_n^{q,a} \big] & = &
\Big[ m(q-1)-n(p-1)\Big]\,
\hat{\psi}_{m+n}^{p+q-2,a}. 
\label{fourcomm}
\eea
Therefore, we obtain all seven commutator relations
given by (\ref{softalgebra}) and (\ref{fourcomm}).
We have checked that the graded Jacobi identities
containing commutator or anticommutator between four currents
are satisfied by using the definition of
$(-1)^{A C} \, \big[ X^A, \big[ X^B, X^C \big\} \big\}
+ \mbox{cycl. perm.}=0$ where $X^A$ denotes a current.
The factor $(-1)^{AC}$ gives us $-1$ for the fermionic currents
$X^A$ and $X^C$ and $1$ for the other three cases.
Subsequently, 
we will describe the corresponding supersymmetric
$w_{1+\infty}$ algebra
in the conventional conformal field theory
\footnote{Three vanishing
anticommutator relations, $\big\{ \hat{\psi}_m^{p,a},
\hat{\psi}_n^{q,b} \big\}=0$, $\big\{ \hat{\psi}_m^{p,a},
\hat{G}_n^{q} \big\}=0$, and $\big\{ \hat{G}_m^{p},
\hat{G}_n^{q} \big\}=0$ from the corresponding regular OPEs in
\cite{FSTZ} must be considered.}.

\section{ A supersymmetric $w_{1+\infty}$ symmetry}

We describe the supersymmetric $w_{1+\infty}$ algebra in order to
understand the symmetry associated with the supersymmetric Einstein-Yang-Mills
theory discussed in the previous section.

\subsection{ A $w_{1+\infty}$ algebra with $SU(N)$ symmetry
}

Odake and Sano \cite{OS}  introduced the affine current $J^{q,a}$
with level $k$, which has weight $q=1, 2, \cdots $
and an adjoint index $a=1,2, \cdots, (N^2-1)$ of $SU(N)$
in addition to the current $w^p$ where  $p=1, 2, \cdots$, into the
$W_{1+\infty}$ algebra \cite{PRS1990-2}.
The results revealed that
the commutators between the currents are determined as follows
\footnote{Previously, the weight was represented by
$p+2$ (or $p+\frac{3}{2}$) rather than an arbitrary $p$ and
we account for this shift properly everywhere.}:
\bea
\big[ W_m^p, W_n^q \big] & = &
\sum_{r\geq 2,\mbox{even}}^{p+q-1} \, \la^{r-2}\, g_{r-2}^{p-2, q-2}(m,n)\,
W_{m+n}^{p+q-r} +
\de^{p q}\, \de_{m+n,0} \, \la^{2 (p-2)}\, c_{p-2}(m), 
\nonu \\
\big[ W_m^p, J_n^{q,a} \big] & = & \sum_{r \geq 2, \mbox{even}}^{p+q-1} \,
\la^{r-2}\, g_{r-2}^{p-2, q-2}(m,n)
\,
J_{m+n}^{p+q-r, a},
\nonu \\
\big[ J_m^{p,a}, J_n^{q,b} \big] & = & \frac{i}{2} \, f^{a b}_{\,\,\,\,c}
\, \sum_{r \geq 1, \mbox{odd}}^{p+q-1} \, \la^{r-2}\, g_{r-2}^{p-2, q-2}(m,n)\,
J_{m+n}^{p+q-r,c} + \de^{p q}\, \de^{a b} \, \de_{m+n,0} \,
\la^{2 (p-2)}\, k_{p-2}(m) \nonu \\
& + &
\sum_{r \geq 2, \mbox{even}}^{p+q-1} \, \la^{r-2}\,
g_{r-2}^{p-2, q-2}(m,n)\, \Big( d^{a b}_{\,\,\,\,c}
J_{m+n}^{p+q-r,c} + \frac{1}{N} \, \de^{a b}\,  W_{m+n}^{p+q-r} \Big).
\label{softalgebra1}
\eea
This is referred to as the $\hat{SU}(N)_k$ $W_{1+\infty}$ algebra
\footnote{
The relations (\ref{softalgebra1})
correspond to equations $(5)$, $(6)$, and $(7)$ of \cite{OS}.
Their $V^{i-2}$ and $W^{j-2,a}$ correspond to
our $W^i$ and $J^{i,a}$, respectively, in this paper.}. The dummy variable $r$ is even or odd \footnote{
In that work, they take the first relation in (\ref{softalgebra1})
from \cite{PRS1990-2} and make an ansatz for the remaining two
with arbitrary coefficients, which can be determined using various Jacobi
identities.}.

By taking the new currents as 
\bea
W_m^p \rightarrow   w_m^p,
\qquad
J_m^{p,a} \rightarrow  \la \, J_m^{p,a},
\label{limit}
\eea
and taking the limit $\la \rightarrow 0$,
the resulting algebra from (\ref{softalgebra1}) can be described as 
\bea
\big[ w_m^p, w_n^q \big] & = &
g_0^{p-2, q-2}(m,n)\, w_{m+n}^{p+q-2}
+ \de^{p, 2} \, \de^{q, 2} \, \de_{m+n,0} \, c_0(m), 
\nonu \\
\big[ w_m^p, J_n^{q,a} \big] & = &  g_0^{p-2, q-2}(m,n)\,
J_{m+n}^{p+q-2, a},
\nonu \\
\big[ J_m^{p,a}, J_n^{q,b} \big] & = & \frac{i}{2} \, f^{a b}_{\,\,\,\,c}
\,  g_{-1}^{p-2, q-2}(m,n)\, J_{m+n}^{p+q-1,c}
+ \de^{p, 1}\, \de^{q, 1} \, \de_{m+n,0} \, k_{-1}(m),
\label{osalgebra}
\eea
where the structure constants are
$g_0^{p-2, q-2}(m,n)= m(q-1)-n(p-1)$ and
$ g_{-1}^{p-2, q-2}(m,n)=\frac{1}{2}$. In addition, the
central terms are given by $c_0(m)=\frac{1}{12} \, m\, (m^2-1) \, c
$ and $ k_{-1}(m)=\frac{m}{16} \, k$.
Note that the central charge $c$ is given by $c =N k$.
Further details are provided in \cite{Sezgin1992}.
Consider $k=0$ and rescaling the current $J^{q,a}$ with
(assuming that the structure constants 
are the same as those mentioned in the  previous section)
the $-\frac{1}{4}$ factor. In this case, the above algebra (\ref{osalgebra})
with wedge modes 
\bea
\big[ w_m^p, w_n^q \big] & = &
\Big[ m(q-1)-n(p-1)\Big]\, w_{m+n}^{p+q-2}, 
\nonu \\
\big[ J_m^{p,a}, J_n^{q,b} \big] & = & -i \, f^{a b}_{\,\,
\,\,c}
\,  J_{m+n}^{p+q-1,c}, 
\nonu \\
\big[ w_m^p, J_n^{q,a} \big] & = & \Big[ m(q-1)-n(p-1)\Big
]\,
J_{m+n}^{p+q-2, a},
\label{threecommutators}
\eea
coincides with the one in (\ref{softalgebra})
when hats are included. In general,
no restrictions are imposed on the modes in
(\ref{threecommutators}) and the weights $p$ and
$q$ are positive integers $p, q =1, 2, \cdots $.
Note that in (\ref{softalgebra}), the modes can be half integers.
The three commutators in (\ref{threecommutators})
are equivalent to the following OPEs in the antiholomorphic sector
(by decomposing the usual mode expansions)
\bea
\label{threeope}
w^{p}(\bar{z}_1) \, w^q(\bar{z}_2) & = & \frac{(p+q-2)}{
(\bar{z}_1-\bar{z}_2)^2} \, w^{p+q-2}(\bar{z}_2)+
\frac{(p-1)}{
  (\bar{z}_1-\bar{z}_2)} \, \bar{\pa}
\, w^{p+q-2}(\bar{z}_2)+\cdots, \nonu \\
J^{p,a}(\bar{z}_1) \, J^{q,b}(\bar{z}_2) & = & 
\frac{-i \, f^{a b}_{\,\,\,\,c}}{
  (\bar{z}_1-\bar{z}_2)} \, J^{p+q-1,c}(\bar{z}_2)+\cdots, \nonu \\
w^{p}(\bar{z}_1) \, J^{q,a}(\bar{z}_2) & = & \frac{(p+q-2)}{
(\bar{z}_1-\bar{z}_2)^2} \, J^{p+q-2,a}(\bar{z}_2)+
\frac{(p-1)}{
  (\bar{z}_1-\bar{z}_2)} \, \bar{\pa}
\, J^{p+q-2,a}(\bar{z}_2)+\cdots.
\eea
According to \cite{PRSplb}, through field redefinitions of the currents,
the weight $1$ current in the $W_{1+\infty}$ algebra is decoupled and
the $W_{\infty}$ algebra is generated by the remaining currents.
Moreover, the so-called $w_N$ algebra, which is a truncation of the
$w_{\infty}$ algebra, is introduced in \cite{Li}.

\subsection{A supersymmetric topological $w_{\infty}$ algebra}

The ${\cal N}=2$ supersymmetric $W_{\infty}$ algebra \cite{BPRSS}
can be twisted to provide a topological $W_{\infty}$ algebra.
By taking one of the fermionic generators as the nilpotent
BRST charge, the corresponding non-trivial commutator relations
are obtained as follows:
\bea
\big[ \hat{V}^p_m, \hat{V}_n^q \big] &=& \sum_{l \geq 2}^{p+q-2} \,
\hat{g}^{p-2, q-2}_{l-2}(m,n)\, \hat{V}_{m+n}^{p+q-l},
\nonu \\
\big[ \hat{V}^p_m, G_{n+\frac{1}{2}}^q \big] &=& \sum_{l \geq 2}
^{p+q-\frac{3}{2}} \,
\hat{g}^{p-2, q-2}_{l-2}(m,n)\, G_{m+n+\frac{1}{2}}^{p+q-l}.
\label{topW}
\eea
Here, no central term is considered. 
The weight of $G^q$ is given by $q =\frac{3}{2}, \frac{5}{2}, \cdots $
while the weight of $\hat{V}^p$ is
given by $p =2, 3, \cdots$ where $p=1$ is excluded. 
The bosonic currents $ \hat{V}^p_m$ in (\ref{topW})
are given by two kinds of bosonic currents
in \cite{BPRSS} \footnote{ The explicit relation can be found in equation
$(8)$ of \cite{PRSS}.}.
By construction,
the bosonic current $\hat{V}^p$ is obtained through linear combination of the
bosonic current $V^p$ of $W_{\infty}$ algebra and the
bosonic current $\tilde{V}^p$ of $W_{1+\infty}$ algebra. 
\footnote{Furthermore, the structure constants
$\hat{g}^{p-2, q-2}_{l-2}(m,n)$
appearing in
(\ref{topW}) are given by equation $(11)$ reported in \cite{PRSS}.}. 
Note that the structure constants in two commutators
are the same.

As in the case of \cite{PRSS}, the relevant contraction
is described by introducing
$v_m^p \rightarrow \la^{p-2} \, \hat{V}_m^p$ and $G^p_m \rightarrow
\la^{p-2} \, G^p_m$ and taking the limit $\la \rightarrow 0$ along the line of
(\ref{limit}) \footnote{Their $\hat{v}^i$ and $g^j$ correspond to
our $v^{i-2}$ and $G^{j-2}$ in this paper.}.
Moreover,
\bea
\big[ v^p_m, v_n^q \big] &=& 
\hat{g}^{p-2, q-2}_0(m,n)\, v_{m+n}^{p+q-2} = \Big[m(q-1)-n(p-1)\Big]\,
 v_{m+n}^{p+q-2},
\nonu \\
\big[ v^p_m, G_{n+\frac{1}{2}}^q \big] &=& 
\hat{g}^{p-2, q-2}_{0}(m,n)\, G_{m+n+\frac{1}{2}}^{p+q-2} =
 \Big[m(q-1)-n(p-1)\Big]\,  G_{m+n+\frac{1}{2}}^{p+q-2},
\label{topalgebra}
\eea
where the non-trivial structure constant
in (\ref{topalgebra})
can be obtained
\footnote{By calculating the four terms in equation
  $(11)$ of \cite{PRSS}, we obtain (\ref{topalgebra}) which 
  occurs in equation $(42)$ in \cite{PRSS}.}
with the help of some formulas in
\cite{BPRSS}
\bea
\hat{g}^{p-2, q-2}_{0}(m,n)& = & \frac{(2 p q-3 p+2 q^2-8 q+8)
(-2 p n-p+2 q m-3 m+2 n+1)}{
2 (2 q-3) (2 p+2 q-5)}
\nonu \\
& + & \frac{(2 p q-3 p+2 q^2-8 q+7)
(-2 p n-p+2 q m-3 m+2 n+1)}{
2 (2 q-3) (2 p+2 q-5)}
\nonu \\
& - & \frac{2 (p-2) (p+m-1)}{(2p-3)}\, \Big(-\frac{1}{4}\Big)
+\frac{2 (p-1) (p+m-1)}{(2p-3)} \, \Big(\frac{1}{4}\Big)
\nonu \\
&=& m(q-1)-n(p-1).
\label{struct}
\eea
Under the $v^2$, the weights of $v^p$ and $G^q$
are $p=2, 3, 4, \cdots $ and $q=2, 3, 4, \cdots$
respectively. During the twisting procedure,
the original weights
of $q$ in the current $G^q$ is shifted by $\frac{1}{2}$.
In (\ref{topalgebra}), the mode of the fermionic current
is given by the half integers (NS sector).
This can be seen from the relation
(\ref{hatone1}) by taking $(n+\frac{1}{2})$
rather than $n$ in the left-hand side.
Further details are provided in \cite{Sezgin1992}. 
In terms of the OPEs,
the following relations corresponding to (\ref{topalgebra})
and (\ref{struct}) are satisfied, similar to the case of (\ref{threeope}),
\bea
\label{OPEtop}
v^{p}(\bar{z}_1) \, v^q(\bar{z}_2)  & = &  \frac{(p+q-2)}{
(\bar{z}_1-\bar{z}_2)^2} \, v^{p+q-2}(\bar{z}_2)+
\frac{(p-1)}{
(\bar{z}_1-\bar{z}_2)} \, \bar{\pa} \, v^{p+q-2}(\bar{z}_2)+\cdots,
\nonu \\
v^{p}(\bar{z}_1) \, G^q(\bar{z}_2) & = & \frac{(p+q-2)}{
(\bar{z}_1-\bar{z}_2)^2} \, G^{p+q-2}(\bar{z}_2)+
\frac{(p-1)}{
(\bar{z}_1-\bar{z}_2)} \, \bar{\pa} \, G^{p+q-2}(\bar{z}_2)+\cdots.  
\eea
From the first equation, we can check the corresponding
commutator relation in
(\ref{topalgebra}). That is,
in the expression of $
\big[ v_m^p, v_n^q \big]=
\oint_{|\bar{z}_1| < \varepsilon}  \, \frac{d \bar{z}_1}{2\pi i}\,
\bar{z}_1^{n+q-1}\,
\oint_{|\bar{z}_{12}| < \varepsilon}  \, \frac{d \bar{z}_2}{2\pi i}\,
\bar{z}_2^{m+p-1}\,
v^p(\bar{z}_1)\, v^q(\bar{z}_2)$,
this leads to $(m+p-1)(p+q-2)-(p-1)(m+n+p+q-2)=m(q-1)-n(p-1)$.
For the second relation of (\ref{OPEtop}) corresponding to
the second relation in
(\ref{topalgebra}), the dummy variable undergoes a shift during the
mode expansion of the fermionic current,  contrary to the case of the bosonic current.

\subsection{ A supersymmetric
$w_{1+\infty}$ algebra with $SU(N)$ symmetry}

By examining the construction of two previous subsections,
we realize that the currents $w_m^p$ in (\ref{osalgebra})
are equivalent to the currents $v_m^p$ in (\ref{topalgebra})
up to $w_m^1$ current \footnote{By decoupling
the $w_m^1$ current with field redefinitions
or inserting the $v_m^1$ current from the analysis performed in
\cite{PRSplb}, we can realize the same number of currents.}.
The currents $\hat{V}_m^p$ in (\ref{topW}) consist of
four parts and two of these parts have no singular OPEs with $J_n^{q,a}$
in (\ref{softalgebra1}). The remaining two terms
have non-trivial OPEs with $J_n^{q,a}$. One of these OPEs is exactly
the same as the $W_m^p$ and the other is a $W_m^{p-1}$ term.
The $W_m^{p-1}$ term provides no contribution after the  abovementioned contraction
procedure ($W_m^p \rightarrow   \la^{p-2} \, w_m^p$ and
$J_m^{p,a} \rightarrow  \la^p \, J_m^{p,a}$) is performed.

In this subsection, we would like to construct
a supersymmetric
$w_{1+\infty}$ algebra with $SU(N)$ symmetry,
which contains the previous algebras (\ref{threecommutators}) and
(\ref{topalgebra}).
We search for this extended algebra in a minimal manner.
In other words,
we introduce the extra currents minimally.
So far, we have the currents $w_m^p$, $G_m^p$, and $J_m^{p,a}$.
We must also determine the superpartner of $J_m^{p,a}$
($G_m^p$ is the superpartner of $w_m^p$).

Let us consider the construction that yields the OPE between
$J^{p,a}$ and $G^q$. We expect that 
a fermionic current with an adjoint index $a$
of $SU(N)$ will occur in the right-hand side of this
OPE.
In other words, the right-hand side of this OPE should
contain the superpartner of $J^{p,a}$ having an adjoint index
$a$. We have previously considered the OPE structure
in the context of $w_{\infty}$ algebra.
That is, 
the OPE between the primary current with weight $p$ and the
current with weight $q$
consists of both the second-order pole and the first-order pole.
The relative coefficients between these are completely fixed
We must now consider one unknown structure constant appearing in the
second-order pole.

Subsequently, we express the following ansatz,
in the antiholomorphic sector,
\bea
J^{p,a}(\bar{z}_1) \, G^q(\bar{z}_2) = \frac{(p+q-2)}{
(\bar{z}_1-\bar{z}_2)^2} \, \psi^{p+q-2,a}(\bar{z}_2)+
\frac{(p-1)}{
(\bar{z}_1-\bar{z}_2)} \, \bar{\pa} \, \psi^{p+q-2,a}(\bar{z}_2)+\cdots.
\label{JGope}
\eea
The coefficient appearing in the second-order pole can be
taken from the normalizations in \cite{PRS1990-1,BK}.
Furthermore, we note that the relative coefficient
$\frac{(p-1)}{(p+q-2)}$ can be determined from the formula
$\frac{\bar{h}_p-\bar{h}_q+\bar{h}_{p+q-2}}{2 \, \bar{h}_{p+q-2}}$
with each weight
$\bar{h}_p=p$, $\bar{h}_q=q$ and $\bar{h}_{p+q-2}=(p+q-2)$
\footnote{
If we assume additional currents, then these currents will appear in other singular terms
up to the central term at the $(p+q)$-th order pole.
The expression for the relative coefficients
can be found in \cite{Blumenhagenetal,Ahn1211}}.
We obtain the corresponding commutator relation
from (\ref{JGope}), by using the procedure reported
in (\ref{OPEtop}),
That is,
\bea
\big[ J_m^{p,a}, G_n^{q} \big] & = &
\Big[ m(q-1)-n(p-1)\Big]\,
\psi_{m+n}^{p+q-2,a}.
\label{commone}
\eea

Subsequently, we obtain the OPEs between
$\psi^{q,b}$
and its superpartner $J_m^{p,a}$ and the currents
$w_m^p$.
From the OPE between the affine currents, we generalize this to
the following OPE that lacks a central term
\bea
J^{p,a}(\bar{z}_1) \, \psi^{q,b}(\bar{z}_2) = 
\frac{-i \, f^{a b}_{\,\,\,\,c}}{
(\bar{z}_1-\bar{z}_2)} \, \psi^{p+q-1,c}(\bar{z}_2)+\cdots.
\label{Jpsiope}
\eea
In terms of the commutator,
we obtain
\bea
\big[ J_m^{p,a}, \psi_n^{q,b} \big] & = &
- i \, f^{a b }_{\,\,\,\,c}\,
\psi^{p+q-1,c}_{m+n}. 
\label{commtwo}
\eea
As usual, from the Jacobi identities between the
generalized affine (bosonic and fermionic) currents,
the sign of the right-hand side of (\ref{commtwo})
can be fixed by using the Jacobi identity between the
structure constants.

For the final OPE we would like to construct,
we expect that the OPE obtained will be similar to the OPE reported in (\ref{JGope}).
The right-hand side of the OPE between
$w^p(\bar{z}_1) \, \psi^{q,a}(\bar{z}_2)$
should contain the
fermionic current having an adjoint index $a$.
Then we obtain the following OPE
\bea
w^p(\bar{z}_1) \, \psi^{q,a}(\bar{z}_2) = \frac{(p+q-2)}{
(\bar{z}_1-\bar{z}_2)^2} \, \psi^{p+q-2,a}(\bar{z}_2)+
\frac{(p-1)}{
  (\bar{z}_1-\bar{z}_2)} \, \bar{\pa} \,
\psi^{p+q-2,a}(\bar{z}_2)+\cdots.
\label{wpsiope}
\eea
The corresponding commutator relation, obtained by following
the procedure reported in (\ref{OPEtop}), can be expressed as follows:
\bea
\big[ w_m^{p}, \psi_n^{q,a} \big] & = &
\Big[ m(q-1)-n(p-1)\Big]\,
\psi_{m+n}^{p+q-2,a}.
\label{commthree}
\eea
This is a natural generalization
in the sense that the fermionic
current $\psi^{q,a}$ is the primary
weight $q$ under the stress energy tensor $w^2$.

Therefore, four additional OPEs, (\ref{JGope}),
(\ref{Jpsiope}), (\ref{wpsiope}) and the second OPE in 
(\ref{OPEtop}) corresponding to (\ref{commone}), (\ref{commtwo}),
(\ref{commthree}), and the second relation of (\ref{topalgebra}), must be considered. 
They, under the wedge modes, 
correspond to the ones in (\ref{fourcomm}) when hats are included.
In this correspondence, we assume
that the weight of the current $w^p$
should be generalized to include the $p=1$ case.
We can determine whether the graded Jacobi identities between four currents
are satisfied
\footnote{ Three vanishing
anticommutator relations are also considered. These are $\big\{ \psi_m^{p,a},
\psi_n^{q,b} \big\}=0$ associated with
vanishing of the central term in the generalization of
the affine current algebra, $\big\{ G_m^{p},
G_n^{q} \big\}=0$ from the result of topological $w_{\infty}$
algebra, and $\big\{ \psi_m^{p,a},
G_n^{q} \big\}=0$.
One realization for the algebra in this paper is given by
the following expressions $w_m^p=i \, y^{p-2} \, e^{i m x} \,
\big[ (p-1) \frac{\pa}{\pa x} - i \, m \, y \, \frac{\pa}{\pa y}
\big]$, $J_m^{q, a}= - i \, t^a \, y^{q-1}\, e^{i m x}$,
$G_m^p=\theta \,w_m^p$
and $\psi_m^{q,a}=\theta \, J_m^{q, a}$ where there are no
$\frac{\pa}{\pa \theta}$ terms from the analysis in
\cite{Sezgin1992}.
}.

\subsection{Appearance of
a supersymmetric $w_{1+\infty}$ symmetry in the context of the celestial conformal
field theory
}

Then we can compare the
symmetry involved in the supersymmetric Einstein-Yang-Mills theory
with
the symmetry we described in the context of the conventional conform
field theory, by using the following field correspondences,
\bea
\hat{w}^p \leftrightarrow w^p, \qquad
\hat{J}^{p,a} \leftrightarrow J^{p,a}, \qquad
\hat{G}^p \leftrightarrow G^p, \qquad
\hat{\psi}^{p,a} \leftrightarrow \psi^{p,a}.
\label{finalequation}
\eea
In (\ref{finalequation}),
the hatted currents have the $\frac{1}{z}$ terms in the
holomorphic and antiholomorphic mode decomposition as in (\ref{H})
and (\ref{modeH}) while the unhatted ones have the standard
antiholomorphic mode decomposition with unrestricted modes.
The weights in the hatted currents
are given by positive integers or half integers
while the weights in the unhatted currents
are given by positive integers.
After we impose the wedge modes onto
the unhatted currents, we observe that
the supersymmetric $w_{1+\infty}$ symmetry
with $SU(N)$ can be described by the celestial conformal
field theory.

\section{ Conclusions and outlook}

We have identified the soft current algebra involved in the
supersymmetric Einstein-Yang-Mills theory with
the supersymmetric $w_{1+\infty}$ algebra where the subalgebra
involves contraction of the i) $\hat{SU}(N)_k$ $W_{1+\infty}$
algebra and ii) topological $W_{\infty}$
algebra. Moreover, three additional OPEs (or
commutator relations) are required for realizing the above soft algebra.

In this paper, we have considered the supersymmetric $w_{1+\infty}$
algebra in an abstract manner without presenting any (free field)
realization. We expect that the bosonic $w_{1+\infty}$ algebra
can be generalized to the $W_{1+\infty}$ algebra at the quantum level,
and hence the quantum version of
the supersymmetric $w_{1+\infty}$
algebra we have obtained must be determined. In other words,
two subalgebras, with quantum
versions that are known before we consider contractions, occur in our findings.
Determining whether the full quantum version of
the (classical)  supersymmetric $w_{1+\infty}$
algebra exists is an open problem. Description of this algebra by
the ${\cal N}=1$ supersymmetric theory may be challenging. In addition, consideration of
the ${\cal N}=2$ supersymmetric $W_{\infty}$ algebra and
addition of the bosonic and fermionic affine currents with the help of the twisting procedure may be manageable. 
In the final check, the Jacobi identity will be used to fix the unknown
structure constants.

Consider the case where the conformal weights of the right-hand side
of the OPE are less than or equal to the
sum of the conformal weights of the left-hand side of the OPE. In this case, the OPE between the currents in the
$W_{\infty}$ algebra contains other high-spin currents.

In other words, in this work, we have identified OPEs
where the non-trivial singular terms are given by both the second- and
the first-order poles in the antiholomorphic sector.
We still face the challenge of identifying all the currents
appearing in the singular terms higher than the third-order pole
associated with the corresponding (supersymmetric) Einstein-Yang-Mills theory.

So far, we have considered the ${\cal N}=1$ supersymmetric
Einstein-Yang-Mills theory. Examining the ${\cal N}=2$ version of this theory may also be an open problem.
Are there any OPEs between the ${\cal N}=2$ soft currents
in the celestial conformal field theory
which correspond to the known ${\cal N}=2$ supersymmetric
$W_{\infty}$ algebra (and its variants)? 
However,
 A ${\cal N}=4$
supersymmetric high-spin algebra exists, as mentioned in the
Introduction, and determining
 whether this algebra is described by the
corresponding celestial conformal field theory 
is an open problem.

\vspace{.7cm}

\centerline{\bf Acknowledgments}

We
would like to
thank
E. Himwich for work on equation $(2.9)$ of \cite{GHPS},
T.R. Taylor for work on the regular OPEs in \cite{FSTZ}
and for the discussions. In addition, we
would like to
thank the referee for various suggestions,
which helped to improve this draft significantly.
This work was supported by
a National Research Foundation of Korea (NRF) grant
funded by the Korean government (MSIT)(No. 2020R1F1A1066893).

\newpage

\appendix

\renewcommand{\theequation}{\Alph{section}\mbox{.}\arabic{equation}}

\section{ A soft current algebra I}

For convenience, we present the result of \cite{GHPS} as follows
\footnote{
\label{iden}
We present two identities
on the finite sums
$\sum_{s=-m-\frac{k}{2}}^{1-k}\, \frac{(-1)^s \, (s+1)\,
(2-s-k-l)!}{(1-s-k)!\, (s+m+\frac{k}{2})!\, (-m-n-
\frac{k+l-2}{2}-s)!} =
\Big[ -m(2-l)+n(2-k)\Big]
\,
\frac{(1-l)!\,
(\frac{2-k}{2} +m +\frac{2-l}{2} +n-1 )!}
{(-1)^{m+\frac{k}{2}}\,
(\frac{2-l}{2} -n)! \,
(\frac{2-k}{2} +m)! \, (\frac{2-l}{2} +n)!}$ and
$\sum_{s=-m+\frac{1-k}{2}}^{1-k}\, \frac{(-1)^s \,
(2-s-k-l)!}{(1-s-k)!\, (s+m+\frac{k-1}{2})!\,
(\frac{1-k}{2}-m+\frac{1-l}{2}-n-s)!}=
\frac{(1-l)!\, 
(\frac{1-k}{2} +m +\frac{1-l}{2} +n )!}{(-1)^{m+\frac{k-1}{2}}\,
(\frac{1-l}{2} -n)! \, (\frac{1-k}{2} +m)! \,
  (\frac{1-l}{2} +n)!}$, which are used in \cite{GHPS}.
We can check, for example,
the first identity inside a mathematica (after introducing
the function $\tt{function[k1_{-},l1_{-},m1_{-},n1_{-}]}$
as a difference between
the left hand and the right hand of identity) as follows:
$\tt{result}=\tt{Table}[0,\{k1,-10,1 \},\{l1,-10,1 \},
  \{m1,-10,10 \},\{n1,-10,10 \}]$.
After that we perform
$\tt{Do[result[[k1,l1,m1,n1]]=FullSimplify[function[k1,l1,m1,n1]];
    Print[``result[``,k1,l1,m1,n1,``]==``}$,
$\tt{result[[k1,l1,m1,n1]]],
\{k1,-3,-1 \},\{l1,-3,-1 \},
  \{m1,1,3 \},\{n1,1,3 \}];  }$.
Then we obtain the zeros for relevant and allowed
$\tt{k1,l1,m1}$ and $\tt{n1}$.}
:
\bea
\big[\hat{R}_m^{k,a}, \hat{R}_n^{l,b} \big] & = & - i \, f^{ab}_{\,\,\,\,\,c}\,
\frac{(\frac{1-k}{2} -m +\frac{1-l}{2} -n )!\,
(\frac{1-k}{2} +m +\frac{1-l}{2} +n )!}{(\frac{1-k}{2} -m)!\,
(\frac{1-l}{2} -n)! \, (\frac{1-k}{2} +m)! \, (\frac{1-l}{2} +n)!}
\, \hat{R}_{m+n}^{k+l-1,c},
\nonu \\
\big[\hat{H}_m^{k}, \hat{H}_n^{l} \big] & = & \frac{\kappa}{2}\,
\Big[ m(2-l)-n(2-k)\Big] \nonu \\
& \times &  
\frac{(\frac{2-k}{2} -m +\frac{2-l}{2} -n-1 )!\,
(\frac{2-k}{2} +m +\frac{2-l}{2} +n-1 )!}{(\frac{2-k}{2} -m)!\,
(\frac{2-l}{2} -n)! \, (\frac{2-k}{2} +m)! \, (\frac{2-l}{2} +n)!}
\, \hat{H}_{m+n}^{k+l},
\nonu \\
\big[\hat{H}_m^{k}, \hat{R}_n^{l,a} \big] & = & \frac{\kappa}{2}\,
\Big[ m(1-l)-n(2-k)\Big] \nonu \\
&\times &  
\frac{(\frac{2-k}{2} -m +\frac{1-l}{2} -n-1 )!\,
(\frac{2-k}{2} +m +\frac{1-l}{2} +n-1 )!}{(\frac{2-k}{2} -m)!\,
(\frac{1-l}{2} -n)! \, (\frac{2-k}{2} +m)! \, (\frac{1-l}{2} +n)!}
\, \hat{R}_{m+n}^{k+l,a}.
\label{GHPSresult}
\eea

\section{ A soft current algebra II}

The four additional commutator relations appearing in
(\ref{rlcomm}), (\ref{hicomm}), (\ref{ricomm}), and (\ref{hlcomm}),
are summarized as follows:
\bea
\big[\hat{R}_m^{k,a}, \hat{L}_n^{l,b} \big] & = & - i \, f^{ab}_{\,\,\,\,\,c}\,
\frac{(\frac{1-k}{2} -m +\frac{\frac{1}{2}-l}{2} -n )!\,
(\frac{1-k}{2} +m +\frac{\frac{1}{2}-l}{2} +n )!}{(\frac{1-k}{2} -m)!\,
(\frac{\frac{1}{2}-l}{2} -n)! \, (\frac{1-k}{2} +m)! \,
(\frac{\frac{1}{2}-l}{2} +n)!}
\, \hat{L}_{m+n}^{k+l-1,c},
\nonu \\
\big[\hat{H}_m^{k}, \hat{I}_n^{l} \big] & = & \frac{\kappa}{2}\,
\Big[ m(\frac{3}{2}-l)-n(2-k)\Big] \nonu \\
& \times &  
\frac{(\frac{2-k}{2} -m +\frac{\frac{3}{2}-l}{2} -n-1 )!\,
(\frac{2-k}{2} +m +\frac{\frac{3}{2}-l}{2} +n-1 )!}
{(\frac{2-k}{2} -m)!\,
(\frac{\frac{3}{2}-l}{2} -n)! \,
(\frac{2-k}{2} +m)! \, (\frac{\frac{3}{2}-l}{2} +n)!}
\, \hat{I}_{m+n}^{k+l},
\nonu \\
\big[\hat{R}_m^{k,a}, \hat{I}_n^{l} \big] & = & \frac{\kappa}{2}\,
\Big[ m(\frac{3}{2}-l)-n(1-k)\Big] \nonu \\
&\times &  
\frac{(\frac{1-k}{2} -m +\frac{\frac{3}{2}-l}{2} -n-1 )!\,
(\frac{1-k}{2} +m +\frac{\frac{3}{2}-l}{2} +n-1 )!}{(\frac{1-k}{2} -m)!\,
(\frac{\frac{3}{2}-l}{2} -n)! \, (\frac{1-k}{2} +m)! \,
(\frac{\frac{3}{2}-l}{2} +n)!}
\, \hat{L}_{m+n}^{k+l,a},
\nonu \\
\big[\hat{H}_m^{k}, \hat{L}_n^{l,a} \big] & = & \frac{\kappa}{2}\,
\Big[ m(\frac{1}{2}-l)-n(2-k)\Big] \nonu \\
&\times &  
\frac{(\frac{2-k}{2} -m +\frac{\frac{1}{2}-l}{2} -n-1 )!\,
(\frac{2-k}{2} +m +\frac{\frac{1}{2}-l}{2}
+n-1 )!}{(\frac{2-k}{2} -m)!\,
(\frac{\frac{1}{2}-l}{2} -n)! \,
(\frac{2-k}{2} +m)! \, (\frac{\frac{1}{2}-l}{2} +n)!}
\, \hat{L}_{m+n}^{k+l,a}.
\label{theresult}
\eea
Here, we present the non-simplified versions of these expressions. 


\end{document}